\documentclass{article} 
\usepackage{iclr2027_conference,times}

\usepackage{amsmath,amsfonts,bm}

\def\eqref#1{equation~\ref{#1}}

\def\1{\bm{1}}

\DeclareMathAlphabet{\mathsfit}{\encodingdefault}{\sfdefault}{m}{sl}
\SetMathAlphabet{\mathsfit}{bold}{\encodingdefault}{\sfdefault}{bx}{n}

\usepackage{url}
\usepackage[utf8]{inputenc} 
\usepackage[T1]{fontenc}    
\usepackage{hyperref}       
\usepackage{url}            
\usepackage{booktabs}       
\usepackage{graphicx}
\usepackage{subcaption}
\usepackage{wrapfig}
\usepackage{amsfonts}       
\usepackage{amsmath}
\usepackage{amssymb}
\usepackage{mathtools}
\usepackage{amsthm}
\usepackage{nicefrac}       
\usepackage{microtype}      
\usepackage{xcolor}         

\usepackage{graphicx}
\usepackage{enumitem}

\newtheorem{proposition}{Proposition}
\newtheorem{lemma}{Lemma}
\newtheorem{corollary}{Corollary}

\newtheorem{assumption}{Assumption}

\usepackage{algorithm}
\usepackage[table]{xcolor}
\usepackage[utf8]{inputenc} 
\usepackage[T1]{fontenc}    
\usepackage{url}            
\usepackage{booktabs}       
\usepackage{nicefrac}       
\newcommand\independent{\protect\mathpalette{\protect\independenT}{\perp}}
\def\independenT#1#2{\mathrel{\rlap{$#1#2$}\mkern2mu{#1#2}}}
\usepackage[english]{babel}
\usepackage{bm}
\DeclareBoldMathCommand\thetab{\theta}
\DeclareBoldMathCommand\phib{\phi}
\DeclareBoldMathCommand\mub{\mu}
\DeclareBoldMathCommand\taub{\tau}
\DeclareBoldMathCommand\sigmab{\sigma}
\DeclareBoldMathCommand\lambdab{\lambda}
\DeclareBoldMathCommand\alphab{\alpha}
\DeclareBoldMathCommand\gammab{\gamma}
\DeclareBoldMathCommand\epsb{\varepsilon}
\newcommand{\Xb}{\mathbf{X}}
\newcommand{\Sb}{\mathbf{S}}
\newcommand{\Cb}{\mathbf{C}}
\newcommand{\xb}{\mathbf{x}}
\newcommand{\cb}{\mathbf{c}}
\newcommand{\sss}{\mathbf{s}}

\title{Provable Speech Attribute Control  \\ via Latent Independence}

\author{Jonathan Svirsky\\
Department of Computer Science\\
Bar Ilan University\\
\texttt{\{jonathan.svirsky\}@biu.ac.il}
\And
Ofir Lindenbaum \\
Faculty of Engineering \\
Bar Ilan University \\
\texttt{\{ofir.lindenbaum\}@biu.ac.il}
\And
Uri Shaham \\
Department of Computer Science \\
Bar Ilan University \\
\texttt{\{uri.shaham\}@biu.ac.il}
}

\iclrfinalcopy 
\begin{document}

\maketitle

\begin{abstract}
Conditional generation and disentangled representation learning are central to controlled generation across audio, vision, and multimodal domains. However, despite strong empirical progress, particularly in speech style transfer, most existing approaches rely on heuristic objectives and architectural choices, offering limited theoretical understanding of when and why reliable attribute control is achievable. In this work, we develop a formal framework for speech attribute conversion and provide a theoretical analysis of sufficient conditions for exact and consistent transfer. Our analysis focuses on a deterministic autoencoder setting augmented with an independence constraint between the learned latent representation and the controllable attribute. Under explicit population-level assumptions about the data-generating process, we establish guarantees linking reconstruction, independence, and the feasibility of attribute manipulation while preserving task-relevant content. We further show how the theoretical framework translates into practice by proposing a practical voice conversion method that directly implements its core principles. Experimental evaluations on voice and pitch conversion tasks demonstrate the applicability of the theoretical analysis to real-world speech conversion settings and show that the resulting method achieves competitive performance against existing approaches.
\end{abstract}

\section{Introduction}

Understanding and controlling structured variability in complex data, such as speech, is a fundamental goal in machine learning. In many applications, observed signals are governed by multiple underlying factors (e.g., linguistic content, speaker identity, emotional tone), and isolating and controlling these components is crucial for tasks such as personalized speech synthesis, cross-lingual voice cloning, and emotion-aware dialogue systems. For instance, cross-lingual voice conversion systems aim to generate speech in a new language while preserving speaker identity \citep{yang21d_interspeech,zhang19e_interspeech}, while emotion transfer models seek to modify the affective content of speech without altering who is speaking \citep{gao2018voice,zhou2023emotion}. These applications assume that meaningful latent representations, such as content and style, can be reliably recovered and manipulated in a disentangled and stable manner. However, ensuring that these latent variables are both identifiable and robustly recoverable remains a fundamental challenge, especially without direct supervision.

Recent advancements in deep learning have spurred significant interest in autoencoder-based (AE) approaches for transforming specific attributes of speech signals, such as speaker identity, emotion, or linguistic content. Techniques such as voice conversion, where the identity of a speaker is altered while preserving the linguistic content \citep{qian2019autovc,polyak2019attention}, and emotion conversion, which transforms emotional expression without affecting speaker identity \citep{niekerk2022comparison,gao2018voice,zhou2021seen}, exemplify the potential of autoencoders for disentangling and manipulating distinct speech attributes.

Despite these empirical successes, a gap remains in the theoretical understanding of whether the true underlying latent variables can be accurately recovered from the observed data and auxiliary inputs. Specifically, it is unclear under what conditions a trained model ensures that the latent representation inferred by the encoder corresponds to the original unobserved variable that generated the data.

We propose an AE framework for structured variable conversion and provide theoretical guarantees for recovering the task-relevant latent variable under explicit assumptions on the generative process. Our setting is a special case of nonlinear Independent Component Analysis (ICA), where we recover only a single latent component, while the remaining components are known and provided as auxiliary information. Although the recovered component may itself be a nonlinear mixture of multiple real-world factors, we show that it is sufficient for accurately converting the input variable. This relaxation of the complete identifiability requirement, which recovers only the relevant component, enables a more tractable and practical approach, especially when disentangling all underlying factors is unnecessary. Our results thus extend the applicability of nonlinear ICA theory (e.g., \citet{hyvarinen2019nonlinear,pmlr-v108-khemakhem20a}) to focused representation learning in conditional generation and variable conversion tasks, such as those found in speech processing.
el{sec:problem}

Let $\Sb \in \mathbb{S}$ be a latent random variable representing speech content, and let $\Cb_i \in \mathbb{C}_i$ be a collection of observed random variables representing different speech characteristics such as speaker identity, pitch, emotion, and others. We use $\Cb$ to denote the random vector $(\Cb_1,...,\Cb_k)^T$. Let $\mathcal{P}$ be a probability distribution over $\mathbb{S}\times\mathbb{C} $ having density $p$. We assume that $\Sb$ and $\Cb$ are independent, i.e., $p$ is an outer product of the marginal densities $p=p_S \times p_C$.

Let $\Xb$ be an observed variable generated as an invertible function $f$ of $\Sb,\Cb$, i.e., $\Xb=f(\Sb,\Cb)$. We denote by $\mathcal{P}_X$ the pushforward distribution induced by $f$, with the corresponding density $p_X$. Our statistical task is for any (new) realizations $\sss, \cb$ to synthesize samples $\mathbf{x} = f(\mathbf{s}, \mathbf{c})$.

The independence assumption above is a modeling assumption on the underlying content variable $\Sb$ and the chosen controllable attributes $\Cb$, not a claim that arbitrary speech factors are independent in natural corpora. Real speech contains correlations between timbre, pitch range, accent, emotion, and rhythm. Our guarantee therefore applies to settings in which the user-specified condition captures the attribute to be changed, and the remaining content variable is approximately invariant to it. Strongly coupled attributes require either redefining the condition set or accepting that exact conversion is outside the scope of the theorem.

\section{Related Work}

Recent efforts in voice conversion and expressive speech modeling have explored diverse directions, including emotional control, disentangled representations, and lightweight architectures. A prominent line of work relies on conditional flow-matching or prompt-based architectures. For example, \citet{pan25b_interspeech} introduces an encoder--decoder framework for emotional speech conversion in which content is extracted using a pretrained ASR HybridFormer \citep{yang2023hybridformer}, while emotion is represented via embeddings from a pretrained contrastive language--audio model. In this setting, the choice of external representations enforces independence between content and emotion by design, while reconstruction is learned with a conditional flow-matching decoder. Similarly, \citet{zuo2025enhancing} employs discrete pitch tokens and target-speaker prompts for expressive conversion within a flow-matching framework, and \citet{wang25ba_interspeech} adopts a token-based in-context learning strategy for controllable voice conversion. In all these approaches, independence between attributes is implicitly assumed through architectural choices rather than explicitly enforced or analyzed, which limits theoretical insight into when and why conversion is guaranteed to succeed.

Another class of methods builds on vector-quantized autoencoders. For instance, \citet{ICLR2025_9328208f} proposes a VQ-VAE-based conversion model that progressively disentangles timbre, style, and linguistic content via discrete latent representations, while \citet{yao2025stablevc} demonstrates controllable zero-shot conversion using conditional flow matching. Although these approaches align naturally with the encoder--decoder paradigm, discreteness alone does not guarantee statistical independence between the learned content representation and the conditioning attributes, nor does it establish conversion correctness. Moreover, codebook learning introduces additional inductive biases and optimization complexity that are difficult to characterize theoretically.

Another line of work focuses on self-supervised disentanglement without relying on external pretrained models. \citet{cai2025genvc} separates linguistic content from speaker style using large-scale self-supervised training on unlabeled corpora, while \citet{liu2025rt} develops a streaming voice conversion system based on differentiable digital signal processing. While these approaches show strong empirical performance, they rely on large models and extensive data and do not provide formal guarantees linking disentanglement to the correctness of synthesis or conversion. 

In contrast, our approach emphasizes reproducibility, accessibility, and theoretical clarity. We design a lightweight architecture trained on moderate-scale public datasets, providing a practical and analyzable baseline for expressive voice conversion. Our design is inspired by \citet{polyak21_interspeech}, where conversion uses discrete units extracted by pretrained content, pitch, and speaker encoders. We generalize this framework by introducing an end-to-end trainable autoencoder that operates directly on speech units containing both linguistic and acoustic information. The encoder is optimized to predict speech units independent of the conditioning variables, and training is performed jointly with the decoder.

A complementary research direction relies on nearest-neighbor search in the embedding space. \citet{baas2023voice} propose KNN-VC, a few-shot voice conversion method in which embeddings from a target speaker’s reference set guide the conversion process. While effective in low-data regimes, inference complexity grows with reference-set size, and correctness depends implicitly on the structure of the embedding space. Our method can be viewed as distilling the implicit invariances exploited by KNN-VC into an explicit latent variable learned by an autoencoder, thereby eliminating the need for nearest-neighbor search and enabling efficient one-shot or short-utterance conversion. Moreover, we generalize autoencoder-based methods by introducing an independence objective that applies to arbitrary conditioning attributes, such as speaker identity or emotion.

Following this direction, \citet{shan2024phoneme} introduces a Phoneme Hallucinator that generates diversified phonemes from short target-speaker references, improving sample quality but inheriting the reliance on synthesized reference samples and nearest-neighbor search, which increases inference latency. In contrast, our method reduces runtime complexity from quadratic to linear with respect to the number of speech samples. Although KNN-VC avoids training a conversion module, our training procedure remains simple and involves two loss terms: independence and reconstruction. Finally, \citet{wang2025voice} extends the KNN-VC framework by clustering semantically similar representations using 2D structural entropy \citep{huang2025structural} and organizing embeddings into a graph. Unlike these approaches, our framework explicitly links latent structure, independence, and conversion correctness through a unified theoretical analysis.

\section{General Framework}
\label{sec:ae_indep}

We introduce the Independence Conditional Autoencoder (ICAE), an effective non-probabilistic framework that avoids priors and posterior inference. ICAE is trained by jointly optimizing two complementary objectives: (1) accurate reconstruction of the input signal, and (2) encouraging statistical independence between the learned latent representation and a given set of conditioning variables. Formally, the objective is defined as:
\begin{equation}
\min_{\thetab_e \cup \thetab_d} \frac{1}{N}\sum_{\xb, \cb} \left[ \mathcal{R}(\mathbf{x}, d_{\thetab_d}(e_{\thetab_e}(\mathbf{x}), \mathbf{c})) + \lambda \mathcal{I}(\mathbf{c}, e_{\thetab_e}(\mathbf{x})) \right],
\label{eq:general_loss}
\end{equation}
where 
\( e(\mathbf{x}) = \hat{\mathbf{s}} \) is the latent representation produced by the encoder \( e \), and \(d(e(\mathbf{x}), \mathbf{c}) = d(\hat{\mathbf{s}}, \mathbf{c}) =  \hat{\mathbf{x}}\) is the reconstructed sample generated by the decoder \( d \). The model is parametrized by a learnable set of parameters, i.e., $\thetab_e$ for the encoder and $\thetab_d$ for the decoder, such that $\thetab = \thetab_e \cup \thetab_d$. The term \( \mathcal{R}(\mathbf{x}, \hat{\mathbf{x}}) \) captures the reconstruction discrepancy, while \( \mathcal{I}(\mathbf{c}, \hat{\mathbf{s}}) \) quantifies the dependence between the latent representation and the conditioning variables. Minimizing this dependence term encourages, but does not by itself certify, independence in finite data. The trade-off between the two objectives is controlled by a scalar \( \lambda > 0 \). 

Our task can be seen as a simplified variant of nonlinear ICA \citep{hyvarinen2019nonlinear}. In this setting, only a single latent component is unknown and must be recovered, while the remaining components are observed. Recent work by \citet{shaham2022discovery} provides theoretical guarantees for latent component analysis; we extend these results to synthesis. Specifically, we show that exact identification of the latent variable is unnecessary for conversion; it is sufficient to recover it up to an arbitrary invertible transformation. We formally show that this level of recovery meets the requirements of the conversion task.

Our primary objective is to learn a decoder $d$ that approximates the generative function $f$, thereby enabling the synthesis of novel samples $\Xb' \sim p_{X}$. Complementing this, we also train an encoder $e$ that recovers a transformed version of $\Sb$, denoted as $\hat{\Sb} = e(\Xb)$. Together, these mappings unify synthesis and analysis: the decoder generates new data consistent with the underlying distribution, while the encoder ensures latent recovery, guiding and validating the conversion task.

In the following sections, we present a theoretical analysis of the proposed framework, defined by the objective function in (\ref{eq:general_loss}), and show that it provides guarantees for style conversion. We also introduce a specific method that instantiates this framework and demonstrates, both theoretically and empirically, that it effectively performs speaker-identity style conversion. These results strengthen the practical applicability of the proposed theoretical framework for real-world style conversion tasks.

\section{Theoretical Analysis}
\label{sec:theoretical}

\subsection{Ideal Variable Conversion}
\label{sec:main_theory}
For the following claims, we assume we have an ICAE model that achieves perfect reconstruction and zero dependence.
\begin{assumption}[\textbf{Perfect Model Convergence}]
\label{assump:zero}
    For a given trained model that is represented by the map $d(e(\xb),\cb)$:
    \begin{align*}
        \mathcal{R}(\mathbf{x}, \hat{\mathbf{x}})=0\text{, }\mathcal{I}(\mathbf{c}, \hat{\mathbf{s}}) = 0, 
        \forall \xb \in \mathbb{X}\text{, s.t. }\mathbf{x}=f(\mathbf{s},\mathbf{c})
    \end{align*}
\end{assumption}
Assumption~\ref{assump:zero} should be understood as an idealized population-level condition,
corresponding to the limit of infinite data, sufficient model capacity, and exact optimization.
It does not assert that such conditions are achieved during finite-sample training. This perspective is standard in identifiability analyses of nonlinear ICA, where recovery guarantees are derived under population-level objectives (e.g., exact likelihood or contrastive loss minimization in the infinite-data regime; see \citet{hyvarinen2019nonlinear}).
Thus, the theory should be read as a set of sufficient conditions and a design principle for conversion models, rather than as a claim that every real speech pipeline exactly satisfies the assumptions. The reconstruction assumption is relaxed in Section \ref{sec:errobound} to derive an error bound.

\begin{assumption}[\textbf{Finite Discrete Latent Variable}]
\label{assump:discrete}
The latent variable $\Sb$ is discrete and admits a finite set of possible realizations,
$\operatorname{supp}(\Sb)=
\{\sss_1,\ldots,\sss_M\},
M \geq 2$, $P_{\Sb}(\sss_m)>0$.
\end{assumption}

This assumption aligns with common preprocessing steps in speech representation learning, such as clustering or quantizing continuous features. The identifiability result below applies to this finite discrete setting; the continuous latent variant used in experiments is included as an empirical comparison rather than as a case covered by the discrete proof.

\begin{assumption}[\textbf{Asymmetric Latent Prior}]
\label{assmp:nonuniform}
For the latent variable $\Sb$ in the generative model $\Xb=f(\Sb,\Cb)$, the symmetry group $\left\{
g:\mathbb{S}\to\mathbb{S}
\;\middle|\;
g \text{ is invertible},\;
g_{\#}P_{\Sb}=P_{\Sb}
\right\}$ contains only the identity map. Equivalently, in the finite discrete setting of Assumption~\ref{assump:discrete}, all point masses are distinct. We define the corresponding \emph{identifiability margin}
\[
\Delta_p
:=
\min_{i\neq j}|p_i-p_j|>0,
\qquad
p_{\max}:=\max_i p_i.
\]
\end{assumption}

Since $\Sb$ is unobserved, Appendix~\ref{app:speech_units} (Figure~\ref{fig:speeh_units_distance}) examines the empirical distribution of a proxy variable $\Tilde{\Sb} \approx \Sb$ derived from real-world speech data. This analysis suggests that symmetry is unlikely in practice, providing empirical support for the approximate satisfaction of Assumption~\ref{assmp:nonuniform}.

The next lemma shows that the learned latent representation $\hat{\Sb}$ preserves all information about the true content $\Sb$ up to an invertible transformation, ensuring that no information relevant for conversion is lost.
\begin{lemma}
\label{lem:T_invertible}
For a given encoder $e$, there is an invertible map $T:\mathbb{S}\to\hat{\mathbb{S}}$ that such that $T(\Sb)=\hat{\Sb}$. 
\end{lemma}
The proof can be found in Appendix \ref{app:T_invertible_proof}. The next proposition shows that the decoder mimics the unknown variable generation function $f$ composed on top of some unknown invertible map $T: \mathbb{S} \rightarrow \hat{\mathbb{S}}$.

\begin{proposition}
\label{prop:equivalence}
Let $T$ be the transformation from Lemma~\ref{lem:T_invertible}. Then, for every admissible condition $\cb$ and every $\hat{\sss}\in T(\mathbb S)=\operatorname{supp}(\hat{\Sb})$,
\[
d(\hat{\sss},\cb)=f(T^{-1}(\hat{\sss}),\cb).
\]
\end{proposition}
Thus, on the learned latent support, the decoder equals the true generative map composed with the inverse latent reparametrization $T^{-1}$. We provide the proof in Appendix \ref{app:equivalence_proof}. The following corollary is the main consequence of the above results, showing that the model performs exact conditional conversion.

\begin{corollary}
\label{coll:conversion}
Under Assumptions~\ref{assump:zero}, \ref{assump:discrete}, and~\ref{assmp:nonuniform}, the conversion of random variable $\Xb$ is guaranteed, i.e. for a given set of samples $\{(\mathbf{x},\mathbf{c}), (\mathbf{x}',\mathbf{c}')\}$ such that $\mathbf{x} =f (\mathbf{s}, \mathbf{c})$ and $\mathbf{x}'=f(\mathbf{s}, \mathbf{c}')$ it holds that:
$d(e(\mathbf{x}), \mathbf{c}') = \mathbf{x}'$.
\end{corollary}

The corollary holds for both seen and unseen realizations of $\Xb$ within the support covered by the assumptions. It guarantees that preserving the condition-independent variable $\Sb$ up to an invertible transformation $T$, while replacing the specified condition $\Cb$, allows us to manipulate samples $\Xb$: we change the attributes represented in $\Cb$ while preserving the remaining information represented by $\Sb$.

\subsection{Model Convergence Implies low Conversion Error}
\label{sec:errobound}

Section~\ref{sec:main_theory} establishes exact conversion under perfect reconstruction and exact independence. We now show that this guarantee degrades gracefully when both conditions are only approximately satisfied, and derive an
error bound for the conversion task.

\begin{assumption}[\textbf{Uniform Reconstruction Error}]
\label{assmp:reconstruction}
There exists $\epsilon\geq0$ such that
\[
\|d(e(\xb),\cb)-\xb\|_2^2\leq\epsilon,
\qquad
\forall\,\xb=f(\sss,\cb)\in\mathbb X.
\]
\end{assumption}

\begin{assumption}[\textbf{$L$-Lipschitz Decoder}]
\label{assmp:smooth}
There exists $L\geq0$ such that
\[
\|d(\hat{\sss}_1,\cb)-d(\hat{\sss}_2,\cb)\|_2
\leq
L\|\hat{\sss}_1-\hat{\sss}_2\|_2,
\]
for all admissible latent values $\hat{\sss}_1,\hat{\sss}_2$ and
$\cb\in\operatorname{supp}(\Cb)$.
\end{assumption}
For each condition value $\cb\in\operatorname{supp}(\Cb)$, let
$
Q_{\cb}:=P_{\hat{\Sb}\mid\Cb=\cb},
Q:=P_{\hat{\Sb}},
$
denote the conditional and marginal distributions of the learned latent representation.
The next assumption requires four intuitive properties. 
First, different realizations of the latent variable $\Sb$ must remain distinguishable in the observation space under the same condition, as expressed by $\gamma_X>0$. 
Second, the reconstruction error must be sufficiently small relative to this separation, so that two distinct latent realizations cannot be mapped to the same learned representation while still being reconstructed accurately. 
Third, the learned representation should be approximately independent of the conditioning variable: for every condition value $\cb$, the conditional distribution $Q_{\cb}$ should remain close to the marginal latent distribution $Q$, with $\delta$ measuring the largest such deviation. In particular, $\delta=0$ recovers exact independence, $\hat{\Sb}\independent\Cb$. 
Finally, changing the condition while keeping the same underlying latent realization $\sss$ should not move its learned representation by an arbitrarily large amount; $D_{\hat S}$ provides a finite bound on this cross-condition variation. 
Together, these conditions ensure that the encoder preserves distinctions between different values of $\Sb$ while remaining approximately insensitive to changes in $\Cb$, which is the key property required for robust conversion.

\begin{assumption}[\textbf{Approximate Independence and Regularity}]
\label{assmp:latentindep}
There exists $\delta\geq0$ such that
\textbf{(1)} $\gamma_X:=
        \inf_{\substack{
        \cb\in\operatorname{supp}(\Cb)\\
        i\neq j
        }}
        \|f(\sss_i,\cb)-f(\sss_j,\cb)\|_2
        >0$,
 \textbf{(2)} $\epsilon<\frac{\gamma_X^2}{4}$, \textbf{(3)} $\operatorname{TV}(Q_{\cb}, Q):=\frac{1}{2}
\sum_{\hat{\sss}\in\hat{\mathbb S}}
|Q_{\cb}(\hat{\sss})-Q(\hat{\sss})|$ and $\sup_{\cb\in\operatorname{supp}(\Cb)}
\operatorname{TV}(Q_{\cb},Q)
\leq\delta$, 
\textbf{(4)} $D_{\hat S}
:=
\sup_{\substack{
\sss\in\operatorname{supp}(\Sb)\\
\cb,\cb'\in\operatorname{supp}(\Cb)
}}
\|e(f(\sss, {\cb}))-e(f(\sss, {\cb}'))\|_2$ is finite.
\end{assumption}

\begin{proposition}[\textbf{Robust Conversion Error Bound}]
\label{prop:error_bound}
Under Assumptions~\ref{assump:discrete},
\ref{assmp:nonuniform},
\ref{assmp:reconstruction},
\ref{assmp:smooth}, and
\ref{assmp:latentindep}, for every
$\cb,\cb'\in\operatorname{supp}(\Cb)$,
let
$\xb_0=f(\Sb,\cb)$, $\xb'=f(\Sb,\cb')$, $\hat{\xb}'=d(e(\xb_0),\cb')$, 
then the conversion error 
$
\epsilon_{\mathrm{conv}}(\Sb,\cb,\cb')
:=
\|\xb'-\hat{\xb}'\|_2^2
$ is bounded:
\begin{align}
\mathbb E_{\Sb}
\left[
\epsilon_{\mathrm{conv}}
\right]
\leq
2\epsilon
+
\frac{8L^2D_{\hat S}^2p_{\max}}{\Delta_p}\delta.
\label{eq:uniform_conversion_bound}
\end{align}
\end{proposition}

\begin{corollary}[\textbf{Robust Conversion Error Bound for Discrete Latent $\hat \Sb$}]
\label{prop:error_bound_discrete}
Assuming that each $\hat \sss \in \mathbb{R}^K$ is a probability vector such that:
$\sum_{k=1}^{K}\hat{s}_k = 1$, the conversion error from Proposition \ref{prop:error_bound} becomes:
\begin{align}
\mathbb E_{\Sb}
\left[
\epsilon_{\mathrm{conv}}
\right]
\leq
2\epsilon
+
\frac{16L^2p_{\max}}{\Delta_p}\delta.
\label{eq:uniform_conversion_bound}
\end{align}
\end{corollary}

Thus, reconstruction error $\epsilon$ and residual dependence $\delta$ are the two quantities that must be driven toward zero, while $L$, $D_{\hat S}$, and $\Delta_p$ determine how strongly these errors are amplified.

The quantity $\Delta_p$ plays a particularly important role as an \emph{identifiability margin}. It measures the minimum separation between the prior probabilities of distinct latent realizations. When $\Delta_p$ is large, assigning a latent value to a code associated with another value produces a detectable change in the conditional distribution of $\hat{\Sb}$. Approximate distributional independence therefore restricts how often such condition-dependent reassignments can occur, yielding approximate paired invariance. Decoder smoothness then propagates this latent agreement to the output space, producing the conversion bound above. Conversely, as $\Delta_p\to0$, the latent prior becomes increasingly symmetric and the guarantee weakens. Consequently, for fixed $L$, $D_{\hat S}$, $p_{\max}$, and $\Delta_p>0$, training the model towards
$\epsilon\to0, \delta\to0 $
implies $\mathbb E_{\Sb}[\epsilon_{\mathrm{conv}}]\to0$,
recovering the exact-conversion regime of Section~\ref{sec:main_theory}. The detailed proof is provided in Appendix~\ref{app:error_proof}.

Having established theoretical guarantees under our assumptions, we now present a practical method that approximates the independence objective in real-world speech conversion tasks using a tractable empirical dependence penalty.

\section{Practical Instantiation of the Framework}
\label{sec:method}

\subsection{Model Architecture}
\begin{figure}[t]
\centering
    \centering
    \includegraphics[width=0.99\linewidth]{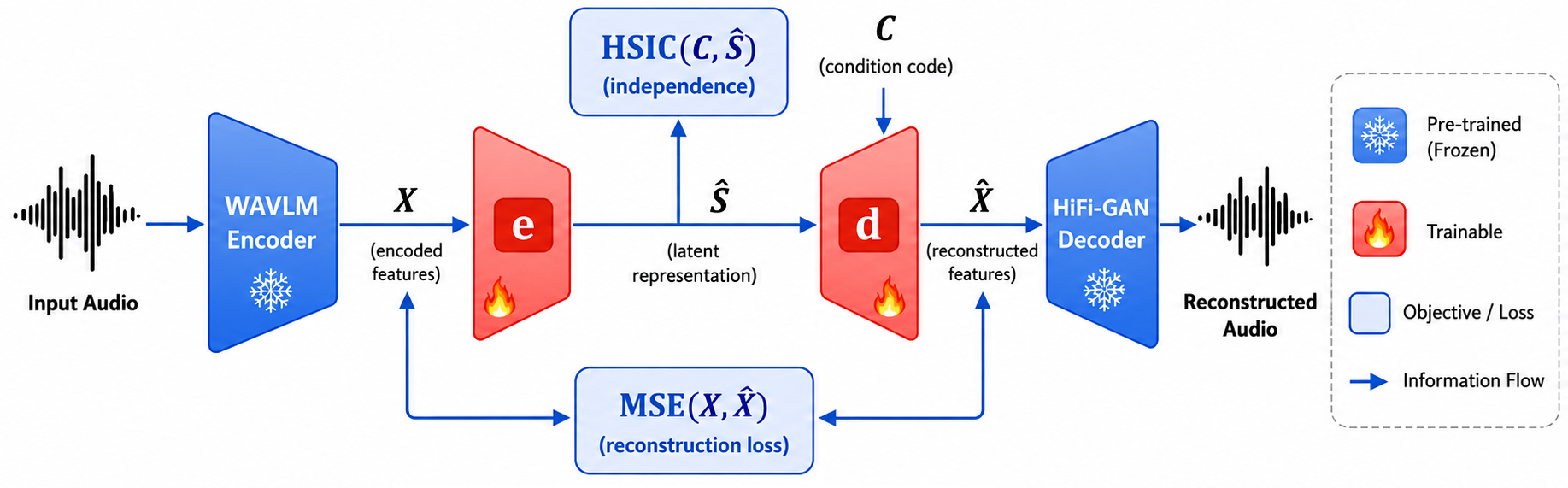}
    \caption{The proposed implementation of the \textbf{IVC} framework for speech attribute conversion. Speech features $\Xb$ are first extracted using a frozen pretrained SSL encoder. The encoder--decoder model $(e,d)$ is then trained to learn a latent representation $\hat{\Sb}$ that is approximately independent of the observed variable $\Cb$, while reconstructing the input features, $\Xb \approx \hat{\Xb}$. Finally, the converted speech waveform is generated using a pretrained acoustic vocoder.}
    \label{fig:cae_indep_model_proposed}
\end{figure}

We next demonstrate how the theoretical framework developed above can be instantiated in a practical speech attribute conversion system. To this end, we introduce the \textbf{Independence-based Voice Conversion (IVC)} method, illustrated in Figure~\ref{fig:cae_indep_model_proposed}. The encoder $e$ maps the input sequence to a shallow latent representation of dimension $D \times T$. The latent representation is  driven by a reconstruction loss and a dependence penalty (HSIC) that encourages statistical independence from the conditioning variable. The decoder $d$ reconstructs the input embeddings.

We focus on the discrete latent representation $\hat{\Sb}$, which directly matches the assumptions used in our main identifiability result. The encoder outputs logits
$ e(\xb)=\hat{\sss}\in\mathbb{R}^{H\times T}$,
which are discretized into one-hot vectors using the Gumbel-Softmax relaxation \citep{jang2017categorical}. The decoder learns an embedding matrix
$
\mathbf{W}\in\mathbb{R}^{K\times H}$,
where $K$ is the dictionary size and $H$ is the dimensionality of each dictionary entry, and reconstructs the input from the embedded discrete representation as
$
\hat{\Xb}=d(\mathbf{W}^{\top}\hat{\Sb},\Cb)$.
For completeness, Appendix~\ref{app:continuous_latent} describes a continuous latent variant and reports its corresponding experimental results.

Both the encoder and decoder are built using non-causal WaveNet residual blocks, as employed in WaveGlow \citep{waveglow2019prenger}, Glow-TTS \citep{NEURIPS2020_5c3b99e8}, and VITS \citep{pmlr-v139-kim21f}. Each WaveNet residual block consists of layers of dilated convolutions, a gated activation unit, and a skip connection. A linear projection layer on top of the residual blocks produces the final sequence of latent embeddings of dimension $H$ for the encoder and reconstructed input representation $\hat{\Xb}$ for the decoder. The decoder receives both the encoder outputs and an additional conditioning tensor. This tensor is first passed through a single convolutional layer and then added before the gated tanh nonlinearities in each residual block \citep{NEURIPS2020_5c3b99e8}. It is formed by concatenating the provided features along the channel dimension, while time-invariant embeddings (e.g., speaker identity) are broadcast along the time axis.

\subsection{Dataset Preparation}

Building on the success of prior works \citep{polyak21_interspeech, baas2023voice, niekerk2022comparison}, we construct our dataset $\mathbb{X}$ from embeddings extracted with a self-supervised learned (SSL) model. Specifically, we employ the WavLM model \citep{chen2022wavlm} to convert waveforms into embeddings taken from its sixth layer, which is known to preserve both semantic and prosodic information \citep{wang2025voice, martin2024investigating, shan2024phoneme, baas2023voice}.
The observable variable $\Cb$ used in this voice-conversion instantiation is an embedding vector for the speaker condition. To extract speaker embeddings, we use a pre-trained speaker encoder, RedimNet \citep{yakovlev24_interspeech}. In addition we experiment with pitch condition which is extracted from Swift-F0 \citep{nieradzik2025swiftf0} as a sequence of continuous pitch values predicted for each time step. 

Finally, to generate waveforms from the converted features, we use a pre-trained acoustic vocoder based on the HiFi-GAN model \citep{kong2020hifi}, pre-trained by \citet{baas2023voice}.

\subsection{Training Objectives}

In this work, we propose a simple and effective objective for encouraging independence in learned representations, inspired by recent advances in discrete speech representation learning \citep{hsu2021hubert, polyak21_interspeech}.

We measure statistical dependence between the latent representation $\hat \Sb$ and the conditioning variable $\Cb$ using the empirical Hilbert--Schmidt Independence Criterion (HSIC). Given a batch of latent sequences $e(\xb) = \hat{\sss} \in \mathbb{R}^{B \times D \times T}$ and corresponding conditions $\cb \in \mathbb{R}^{B \times D_c \times T}$, we first reshape them into a set of $n = B \cdot T$ paired samples:
\(
\{(\hat{\sss}_i, \cb_i)\}_{i=1}^n, \quad \hat{\sss}_i \in \mathbb{R}^D, \; \cb_i \in \mathbb{R}^{D_c}.
\)
For time-invariant conditions such as speaker identity, the same condition vector is broadcast across frames before this reshape. We therefore enforce independence at the frame level by applying the HSIC objective between the framewise latent representations and the corresponding condition. This choice is motivated by the local structure of self-supervised speech representations, where individual frames encode substantial phonetic and acoustic information; indeed, framewise representations have been successfully used for content-preserving voice conversion in prior work such as KNN-VC. 

To capture nonlinear dependence, we compute a biased centered HSIC estimator using radial basis function (RBF) kernels \footnote{We use a multi-scale RBF kernel composed of three kernels with different bandwidths, where each bandwidth is set to the maximum squared distance to the m-th nearest neighbor across all samples $(m=5, 10, 20)$.}. Let $\mathbf{K}_{\hat{\Sb}}
=k(\hat{\sss}), \mathbf{K}_{\hat{\Cb}}=k(\hat{\cb}) \in \mathbb{R}^{n \times n}$ be kernel matrices with zero diagonal, then the dependence loss term is given by:
\begin{equation}
\label{eq:hsic}
    \mathcal{L}_{\mathrm{HSIC}}(\hat{\mathbf{s}}, \mathbf{c})
=(n-1)^{-2}\text{tr}(\mathbf{K}_{\hat{\Sb}} \mathbf{H} \mathbf{K}_{\Cb} \mathbf{H}),
\end{equation}
where $\mathbf{H} \in  \mathbb{R}^{n \times n}$ is the centering matrix defined by $\mathbf{H}=\mathbf{I}_n-\frac{1}{n}\mathbf{1}_n\mathbf{1}_n^T$ to ensure that the kernels are computed on zero-mean features.
Intuitively, HSIC measures the covariance between features in two Reproducing Kernel Hilbert Spaces induced by the kernels on $\hat{\Sb}$ and $\Cb$. If the variables are statistically independent, the cross-covariance operator vanishes, and thus the objective in Eq.~\ref{eq:hsic} approaches zero. Finally, the reconstruction is enforced by the mean-squared error loss term:
\[
\mathcal{L}_{\text{recon}} (\xb, \cb) = \lVert\mathbf{x} -  d(e(\mathbf{x}), \cb) \rVert_2^2
\]
The model is then trained with the following objective:
\begin{align*}
    \mathcal{L}_{\text{\textbf{IVC}}} (\xb, \cb)
     =\frac{1}{N}\sum_{(\mathbf{x}, \cb)} [ \underbrace{ \mathcal{L}_{\text{recon}}}_{\mathcal{R}(\cdot)}(\xb, \cb)  + \lambda \underbrace{\mathcal{L}_{\text{HSIC}}(\hat \sss, \cb)}_{\mathcal{I}(\cdot)} ],
\end{align*}
where $\lambda$ is a hyperparameter. 
A key advantage of this formulation is its simplicity. Unlike other approaches, such as vCLUB \citep{cheng2020club} or discriminator-based methods \citep{shaham2022discovery}, our optimization avoids adversarial training, which is notoriously unstable. 
Moreover, our method supports one-shot inference, requiring only a single reference speaker example at test time. It also achieves a linear inference-time complexity with respect to the number of input samples, since it bypasses pairwise similarity computations and neighbor-search steps used in methods such as KNN-VC.

\section{Experiments}
\label{sec:experiments}
\subsection{Datasets}

We train our model with the VCTK dataset (110 speakers), LibriTTS \citep{zen19_interspeech} train-clean-360 (904 speakers), LibriSpeech train-clean-100 (251 speakers) \citep{panayotov2015librispeech}, and ESD English (10 speakers) \citep{zhou2021seen}.
This dataset yields 1,265 speakers.
To evaluate any-to-any voice conversion, we conduct our experiments on the LibriSpeech test-clean dataset \citep{panayotov2015librispeech}, which comprises 40 speakers. To perform the voice conversion, we constructed 7800 content-reference pairs from the test-clean set of LibriSpeech by taking 5 utterances for each speaker and
converting them into the remaining 39 speakers, which gives a
total of 40 × 5 × 39 = 7800 samples. During inference, the reference speaker is provided using a single 3-second center-cropped utterance. We provide audio samples on  GitHub\footnote{\url{https://ivc2027.github.io/ivc/}}.

\subsection{Evaluation}

We evaluate our method on the voice conversion task by measuring word error rate (WER) and character error rate (CER) for speech intelligibility. For speaker similarity, we report the target-vs-converted equal error rate (Target EER): higher values indicate that the verification model finds converted samples harder to distinguish from target-speaker samples. This is therefore a similarity score derived from an EER protocol, not a conventional same/different speaker verification error where lower is better. For intelligibility, we utilize the Wav2Vec2-Base model \citep{baevski2020wav2vec}, and for speaker similarity, we employ the speaker verification system developed by \citet{snyder2018xvectors} and implemented by \citet{ravanelli2021speechbrain}. For assessing the naturalness of the converted speech, we adopt the experimental protocol introduced in the recent GenVC study~\citep{cai2025genvc} and use the UTMOSv2 model to predict perceptual scores that have been shown to correlate strongly with human mean opinion score (MOS) judgments~\citep{baba2024t05}. Objective evaluation results are reported as UTMOSv2, with all values accompanied by 95\% confidence intervals (higher is better). Evaluation is conducted on the EMIME dataset, which, consistent with~\citep{cai2025genvc}, serves as a test set containing speakers unseen during training for all evaluated voice conversion models. We treat UTMOSv2 as an automatic proxy for naturalness rather than a substitute for a human listening study.

In addition, we evaluate our method on the pitch transfer task. We use the Pearson correlation coefficient (PCC) between the target and generated pitch trajectories. The trajectories are first aligned using dynamic time warping (DTW), after which PCC is computed on the aligned sequences. Emotion labels are used only to define source--target utterance pairs and are not provided as conditioning variables. Additiona details are provided in Appendix~\ref{app:pitch}.
\begin{table}[t]
    \centering
  \caption{Comparison of different methods on objective evaluation metrics. The reference speaker is given by a single \textbf{ 3-second} utterance.}
    \resizebox{0.7\linewidth}{!}{
\begin{tabular}{lccc}
\toprule
Method & WER (\%) $\downarrow$ & CER (\%) $\downarrow$ & EER (\%) $\uparrow$ \\
\midrule
Target  & 3.39 & 0.84 & 47.83 \\
\hline
KNN-VC \citep{baas2023voice}        & 43.84 & 21.38 & 12.12 \\
FreeVC \citep{li2023freevc}        & 4.98  & 1.51  & 9.86  \\
MKL-VC \citep{lobashev2025training}        &\textbf{ 3.76 } & \textbf{1.02}  & 9.12  \\
GENVC-Small   \citep{cai2025genvc}      & 10.20    & 3.81    & 12.97    \\
MeanVC \citep{ma2026meanvc}       & 29.26    & 13.25    & 6.14 \\
\rowcolor[HTML]{dff2a6} \textbf{IVC}   & 5.29  & 1.59  & \textbf{23.00} \\  
\bottomrule
\end{tabular}}
    \label{tab:oneshot}
\end{table}

\begin{table}[t]
\centering
\caption{Objective naturalness evaluation using UTMOSv2 (higher is better) on the EMIME dataset. All results are reported with a 95\% confidence interval.
}
\label{tab:utmos_results}
\resizebox{0.5\linewidth}{!}{
\begin{tabular}{l c}
\toprule
Model & UTMOSv2 $\uparrow$ \\
\midrule
YourTTS \citep{casanova2022yourtts} & $2.43 \pm 0.018$ \\
FreeVC \citep{li2023freevc} & $2.81 \pm 0.014$ \\
NEURAL VC~\citep{cao2024neuralvc} & $2.60 \pm 0.015$ \\
GENVC-SMALL~\citep{cai2025genvc} & $2.66 \pm 0.021$ \\
GENVC-LARGE~\citep{cai2025genvc} & $2.65 \pm 0.019$ \\
\midrule
\rowcolor[HTML]{dff2a6}  \textbf{IVC} & \textbf{3.14} $\pm$ 0.015 \\
\bottomrule
\end{tabular}}
\vskip -0.2in
\end{table}

\subsection{Results}
Table \ref{tab:oneshot} presents the speaker conversion results of the proposed method. All baselines were run by us with the same inference setup as our method. Table \ref{tab:oneshot} shows that our method achieves the strongest Target EER speaker-similarity score under this protocol when a short reference utterance is provided for voice conversion. The proposed method also improves over KNN-VC in intelligibility (WER, CER metrics) and speaker similarity, and it achieves intelligibility results comparable to the MKL-VC \citep{lobashev2025training} and Free-VC \citep{li2023freevc} baselines. The discrete and continuous variants perform similarly on intelligibility, while the discrete variant obtains the strongest Target EER among the compared conversion methods, consistent with the theory's emphasis on discrete latent content. Table~\ref{tab:utmos_results} reports the UTMOSv2 results for our method alongside the models evaluated under the same experimental conditions in~\citep{cai2025genvc}. These automatic metrics suggest that the proposed design is a useful and reproducible baseline, while human listening tests remain necessary for a complete perceptual evaluation. Finally, our method achieves the second-best performance on the pitch conversion task, as shown in Table~\ref{tab:pitch_corr_summary}, while remaining substantially more parameter-efficient. In particular, GENVC uses more than 500M parameters, whereas our full pipeline contains approximately 122M parameters.

\begin{table}[h]
\centering
\caption{Average pitch correlation (PCC) across all source--target emotion directions.}
\vskip -0.1in
\resizebox{0.9\linewidth}{!}{
\begin{tabular}{lcccccc}
\toprule
 & \textbf{MKL-VC} & \textbf{FreeVC} & \textbf{KNN-VC} & \textbf{MeanVC} & \textbf{GENVC} & \colorbox[HTML]{DFF2A6}{\textbf{IVC}} \\
\midrule
PPC across 20 directions & 0.668 & 0.729 & 0.726 & 0.615 & \textbf{0.754} & \underline{0.736} \\
\bottomrule
\end{tabular}}
\vskip -0.1in
\label{tab:pitch_corr_summary}
\end{table}

\section{Conclusion}

We introduced a principled framework for speech attribute conversion with formal guarantees. Our analysis shows that recovering the latent variable $\Sb$ up to an invertible transformation is sufficient for correct synthesis and attribute manipulation, and we further characterize the effect of imperfect training through explicit conversion-error bounds.

Our method is simple and reproducible: it relies only on reconstruction and independence losses, avoids adversarial training, and uses a lightweight architecture for efficient one-shot any-to-any voice conversion. Despite its simplicity, the proposed approach achieves competitive intelligibility and strong performance under the EER-based speaker-similarity protocol, and outperforms all baselines in the pitch conversion task. The guarantees are intentionally stated under explicit idealized assumptions, and the experiments should be read as evidence that the resulting design principle is useful rather than as a claim that the assumptions hold exactly in real speech data. Important remaining limitations include the absence of a large-scale human perceptual study and the reliance on automatic proxies such as UTMOSv2. Overall, this work bridges theory and practice in controllable speech generation, establishing IVC as a reliable baseline and a foundation for future theory-grounded speech conversion research.

\section*{AI use statement}

In this work, we used generative AI tools only to improve the writing style, clarity, and grammar of the manuscript. We did not use generative AI tools to formulate scientific claims, derive theoretical results, construct proofs, analyze experimental results, or draw conclusions. All scientific claims, mathematical derivations, proofs, experimental design, analyses, and conclusions are the authors’ own work.

We have reviewed all AI-assisted edits and verified that they preserve the intended technical meaning and do not introduce unsupported statements or modifications to the scientific content. We take full responsibility for the final content of this work, including any text revised with the aid of generative AI.

\bibliography{refs}
\bibliographystyle{iclr2027_conference}

\appendix
\section{Pitch Conversion Results}
\label{app:pitch}
This experiment evaluates pitch-sequence transfer across emotion-labeled utterance pairs, rather than full emotion conversion. The experimental setup is as follows:
\begin{itemize}
    \item \textbf{Condition $\Cb$:} a time-varying pitch trajectory extracted using Swift-F0
    \item \textbf{Source signal:} an utterance wtih acoustic-content representation
    \item \textbf{Target condition:} the pitch trajectory extracted from an utterance belonging to the target emotion category
    \item \textbf{Output:} speech reconstructed from the source latent representation and the target pitch sequence
    \item \textbf{Evaluation metric:} Pearson correlation coefficient (PCC) between the generated and target pitch trajectories after dynamic-time-warping (DTW) alignment.
\end{itemize}

The emotion labels are used only to define source--target groups, such as Angry~$\rightarrow$~Happy. Emotion itself is not provided as the conditioning variable, and therefore this experiment does not establish full emotion conversion. We accordingly refer to this setting as \emph{pitch transfer across emotion-labeled pairs}.
\begin{table*}[h]
\centering
\caption{Pitch correlation for emotion conversion across source-target emotion directions.}
\resizebox{0.9\linewidth}{!}{

\begin{tabular}{lcccccc}
\toprule
\textbf{Conversion Direction} & \textbf{MKL-VC} & \textbf{FreeVC} & \textbf{KNN-VC} & \textbf{MeanVC} & \textbf{GENVC} & \colorbox[HTML]{DFF2A6}{\textbf{IVC}} \\
\midrule
Angry $\rightarrow$ Happy      & 0.738 & 0.671 & 0.796 & 0.733 & 0.793 & 0.823 \\
Angry $\rightarrow$ Neutral    & 0.789 & 0.899 & 0.700 & 0.635 & 0.790 & 0.875 \\
Angry $\rightarrow$ Sad        & 0.776 & 0.859 & 0.755 & 0.656 & 0.786 & 0.821 \\
Angry $\rightarrow$ Surprise   & 0.600 & 0.564 & 0.702 & 0.726 & 0.718 & 0.730 \\
Happy $\rightarrow$ Angry      & 0.710 & 0.743 & 0.757 & 0.582 & 0.641 & 0.570 \\
Happy $\rightarrow$ Neutral    & 0.763 & 0.916 & 0.732 & 0.676 & 0.814 & 0.782 \\
Happy $\rightarrow$ Sad        & 0.682 & 0.830 & 0.674 & 0.377 & 0.773 & 0.752 \\
Happy $\rightarrow$ Surprise   & 0.665 & 0.645 & 0.692 & 0.686 & 0.637 & 0.662 \\
Neutral $\rightarrow$ Angry    & 0.638 & 0.724 & 0.701 & 0.758 & 0.781 & 0.782 \\
Neutral $\rightarrow$ Happy    & 0.653 & 0.695 & 0.776 & 0.776 & 0.769 & 0.767 \\
Neutral $\rightarrow$ Sad      & 0.628 & 0.831 & 0.686 & 0.833 & 0.803 & 0.808 \\
Neutral $\rightarrow$ Surprise & 0.556 & 0.572 & 0.648 & 0.689 & 0.647 & 0.698 \\
Sad $\rightarrow$ Angry        & 0.584 & 0.659 & 0.768 & 0.570 & 0.695 & 0.617 \\
Sad $\rightarrow$ Happy        & 0.636 & 0.671 & 0.823 & 0.615 & 0.758 & 0.782 \\
Sad $\rightarrow$ Neutral      & 0.684 & 0.873 & 0.722 & 0.730 & 0.822 & 0.686 \\
Sad $\rightarrow$ Surprise     & 0.545 & 0.527 & 0.620 & 0.506 & 0.714 & 0.611 \\
Surprise $\rightarrow$ Angry   & 0.604 & 0.692 & 0.744 & 0.466 & 0.669 & 0.793 \\
Surprise $\rightarrow$ Happy   & 0.788 & 0.820 & 0.822 & 0.708 & 0.794 & 0.799 \\
Surprise $\rightarrow$ Neutral & 0.718 & 0.755 & 0.760 & 0.414 & 0.844 & 0.677 \\
Surprise $\rightarrow$ Sad     & 0.597 & 0.638 & 0.649 & 0.158 & 0.824 & 0.683 \\
\midrule
Average & 0.668 & 0.729 & 0.726 & 0.615 & \textbf{0.754} & \underline{0.736} \\
\bottomrule
\end{tabular}}
\label{tab:pitch_corr_emotion}

\end{table*}

\section{Ablation Study}

\begin{table}[h]
    \centering
  \caption{Evaluation of different implementations for our method. \textbf{IVC}$^{\mathcal{R}}$ is trained without HSIC loss term.}
    \resizebox{0.65\linewidth}{!}{
\begin{tabular}{lccc}
\toprule
Method & WER (\%) $\downarrow$ & CER (\%) $\downarrow$ & EER (\%) $\uparrow$ \\
\midrule
\textbf{IVC}   & 5.18  & 1.58  & 23.00 \\ 

\textbf{IVC}$^{\text{Cont}} (\text{continuous } \hat{\Sb})$   & 5.30  & 1.60  & 20.50 \\

\textbf{IVC}$^{\mathcal{R}}(\text{without HSIC })$   & 5.29  & 1.59  & 22.67 \\
\bottomrule
\end{tabular}}
 \vskip -0.1in
    \label{tab:ablation_ivc}
\end{table}

\section{Additional Lemmas}
For a fixed condition value $\cb \in \mathbb{C}$, we define $\mathbb{X}^c$ as the support of the conditional distribution of $\Xb$ given $\Cb=\cb$, namely
\[
\mathbb{X}^c 
:= \big\{ \xb \in \mathbb{X} \;:\; \xb = f(\sss,\cb), \ \sss \in \mathrm{supp}(\Sb) \big\}
\;\subseteq\; \mathbb{X}.
\]
We further define the encoder map $e^c : \mathbb{X}^c \to \hat{\mathbb{S}}$ by
\(
e^c(\xb) := e(\xb),  \forall \xb \in \mathbb{X}^c
\), and similarly define $d^c : \hat{\mathbb{S}} \to \hat{\mathbb{X}}$ by $d^c(\hat \sss):=d(\hat \sss, \cb)$

\begin{lemma}
\label{lem:e_c_injective}
Under Assumption~\ref{assump:zero}, for every admissible condition $\cb$, the restricted encoder map $e^c:\mathbb{X}^c\rightarrow\hat{\mathbb{S}}$ is injective.
\end{lemma}

\begin{proof}
Fix a condition value $\cb$ and take any $\xb_1,\xb_2\in\mathbb{X}^c$. Suppose that $e^c(\xb_1)=e^c(\xb_2)=\hat{\sss}$. By perfect reconstruction in Assumption~\ref{assump:zero},
\[
\xb_1=d(\hat{\sss},\cb)=\xb_2.
\]
Therefore equality of encoder outputs implies equality of inputs, which is exactly injectivity of $e^c$ on $\mathbb{X}^c$.
\end{proof}

\section{Proof of Lemma \ref{lem:T_invertible}}
\label{app:T_invertible_proof}
\begin{proof}
For every admissible condition $\cb$, define
\[
T^c:\mathbb S\to\hat{\mathbb S},
\qquad
T^c(\sss):=e(f(\sss,\cb)).
\]
Because $f(\cdot,\cb)$ is injective in $\sss$ by the assumed invertibility of the generative map and $e^c$ is injective by Lemma~\ref{lem:e_c_injective}, their composition $T^c=e^c\circ f(\cdot,\cb)$ is injective.

We next show that $T^c$ cannot depend on $\cb$. Fix two admissible condition values $\cb_1,\cb_2$ and suppose, for contradiction, that there is an index $i$ such that
\[
T^{c_1}(\sss_i)\neq T^{c_2}(\sss_i).
\]
Set $\hat{\sss}_i:=T^{c_1}(\sss_i)$. Since $T^{c_1}$ is injective and $P_{\Sb}(\sss_i)=p_i>0$, the conditional distribution of $\hat{\Sb}$ given $\Cb=\cb_1$ assigns mass
\[
P_{\hat{\Sb}\mid\Cb=\cb_1}(\hat{\sss}_i)=p_i.
\]
Now consider the same latent point $\hat{\sss}_i$ under condition $\cb_2$. There are two exhaustive cases.

\emph{Case 1:} there is no $j$ such that $T^{c_2}(\sss_j)=\hat{\sss}_i$. Then
\[
P_{\hat{\Sb}\mid\Cb=\cb_2}(\hat{\sss}_i)=0,
\]
which differs from $p_i>0$.

\emph{Case 2:} there exists $j$ such that $T^{c_2}(\sss_j)=\hat{\sss}_i$. Injectivity of $T^{c_2}$ makes this $j$ unique. Moreover $j\neq i$, because by assumption $T^{c_2}(\sss_i)\neq\hat{\sss}_i$. Hence
\[
P_{\hat{\Sb}\mid\Cb=\cb_2}(\hat{\sss}_i)=p_j.
\]
Assumption~\ref{assmp:nonuniform} gives $p_j\neq p_i$.

Thus in either case the conditional law of $\hat{\Sb}$ differs between $\cb_1$ and $\cb_2$. This contradicts the zero-dependence part of Assumption~\ref{assump:zero}, which implies $\hat{\Sb}\independent\Cb$ and therefore
\[
P_{\hat{\Sb}\mid\Cb=\cb_1}
=
P_{\hat{\Sb}\mid\Cb=\cb_2}
\]
for admissible condition values (equivalently, for $P_{\Cb}$-almost every pair in the general measure-theoretic formulation).

Consequently $T^{c_1}(\sss_i)=T^{c_2}(\sss_i)$ for every $i$ and every pair of admissible conditions. Hence all condition-specific maps coincide with a single map $T$, and
\[
e(f(\sss,\cb))=T(\sss)
\qquad
\forall\,\sss\in\mathbb S,\ \cb\in\operatorname{supp}(\Cb).
\]
Since each $T^c$ is injective, $T$ is injective. Because $\mathbb S$ is finite, $T$ is a bijection from $\mathbb S$ onto its image $T(\mathbb S)$ and therefore has a well-defined inverse on that image. Finally, $\hat{\Sb}=e(\Xb)=e(f(\Sb,\Cb))=T(\Sb)$ almost surely, completing the proof.
\end{proof}

\section{Proof of Proposition \ref{prop:equivalence}}
\label{app:equivalence_proof}
\begin{proof}
Take any $\hat{\sss}\in T(\mathbb S)$. Since $T$ is bijective from $\mathbb S$ onto $T(\mathbb S)$ by Lemma~\ref{lem:T_invertible}, there exists a unique $\sss=T^{-1}(\hat{\sss})$. For any admissible condition $\cb$, Lemma~\ref{lem:T_invertible} gives
\[
e(f(\sss,\cb))=T(\sss)=\hat{\sss}.
\]
Applying perfect reconstruction from Assumption~\ref{assump:zero} to the sample $\xb=f(\sss,\cb)$ yields
\[
d(e(f(\sss,\cb)),\cb)=f(\sss,\cb).
\]
Substituting $e(f(\sss,\cb))=\hat{\sss}$ and $\sss=T^{-1}(\hat{\sss})$ gives
\[
d(\hat{\sss},\cb)
=
f(\sss,\cb)
=
f(T^{-1}(\hat{\sss}),\cb).
\]
Since $\hat{\sss}\in T(\mathbb S)$ and $\cb$ were arbitrary, the equality holds on the entire learned latent support for every admissible condition.
\end{proof}

\section{Proof of Corollary~\ref{coll:conversion}}
\label{app:coll_proof}
\begin{proof}
By applying Proposition \ref{prop:equivalence}: $d(e(\mathbf{x}),\mathbf{c}') = d(\hat{\mathbf{s}}, \mathbf{c}') = f (T^{-1} (\hat{\mathbf{s}}), \mathbf{c}') = f (\mathbf{s}, \mathbf{c}') = \mathbf{x}'$
\end{proof}

\section{Proof of Proposition~\ref{prop:error_bound}}
\label{app:error_proof}

\begin{proof}
Fix any two condition values
$\cb,\cb'\in\operatorname{supp}(\Cb)$.
Recall that
\[
T_{\cb}(\sss):=e(f(\sss,\cb)),
\qquad
Q_{\cb}:=(T_{\cb})_{\#}P_{\Sb}
=
P_{\hat{\Sb}\mid\Cb=\cb}.
\]
We prove the result in four steps.

\paragraph{Step 1: imperfect reconstruction preserves injectivity.}
We first show that $T_{\cb}$ is injective for every
$\cb\in\operatorname{supp}(\Cb)$.
Suppose, for contradiction, that there exist two distinct latent states
$\sss_i\neq\sss_j$ such that
\[
T_{\cb}(\sss_i)=T_{\cb}(\sss_j)=\hat{\sss}.
\]
By definition of $T_{\cb}$,
\[
e(f(\sss_i,\cb))
=
e(f(\sss_j,\cb))
=
\hat{\sss}.
\]
Assumption~\ref{assmp:reconstruction} therefore gives
\(
\|d(\hat{\sss},\cb)-f(\sss_i,\cb)\|_2
\leq \sqrt{\epsilon}
\)
and
\(
\|d(\hat{\sss},\cb)-f(\sss_j,\cb)\|_2
\leq \sqrt{\epsilon}.
\)
Using the triangle inequality,
\begin{align*}
\|f(\sss_i,\cb)-f(\sss_j,\cb)\|_2
&\leq
\|f(\sss_i,\cb)-d(\hat{\sss},\cb)\|_2
+
\|d(\hat{\sss},\cb)-f(\sss_j,\cb)\|_2\\
&\leq 2\sqrt{\epsilon}.
\end{align*}
However, by the definition of $\gamma_X$,
\[
\|f(\sss_i,\cb)-f(\sss_j,\cb)\|_2
\geq \gamma_X.
\]
Assumption~\ref{assmp:latentindep} states that
$\epsilon<\gamma_X^2/4$, and hence
\[
2\sqrt{\epsilon}<\gamma_X,
\]
which gives a contradiction. Therefore, $T_{\cb}$ is injective for every
admissible condition value $\cb$.

\paragraph{Step 2: total variation controls condition-dependent latent disagreement.}
Write
\[
p_i:=P_{\Sb}(\sss_i),
\qquad i=1,\ldots,K,
\]
and recall from Assumption~\ref{assmp:nonuniform} that
\[
\Delta_p
:=
\min_{i\neq j}|p_i-p_j|>0,
\qquad
p_{\max}:=\max_i p_i.
\]

For the fixed pair $\cb,\cb'$, define
\[
\hat{\sss}_i:=T_{\cb}(\sss_i),
\qquad
\hat{\sss}'_i:=T_{\cb'}(\sss_i),
\]
and let
\[
M_{\cb,\cb'}
:=
\left\{
i:
T_{\cb}(\sss_i)\neq T_{\cb'}(\sss_i)
\right\}
\]
be the set of latent states whose learned representation changes between
the two condition values.

Since both $T_{\cb}$ and $T_{\cb'}$ are injective, each point
$\hat{\sss}_i=T_{\cb}(\sss_i)$ has probability
\[
Q_{\cb}(\hat{\sss}_i)=p_i.
\]
Under $Q_{\cb'}$, there are two possibilities.

First, suppose that $\hat{\sss}_i$ does not belong to the image of
$T_{\cb'}$. Then
\(
Q_{\cb'}(\hat{\sss}_i)=0,
\)
and therefore
\[
|Q_{\cb}(\hat{\sss}_i)-Q_{\cb'}(\hat{\sss}_i)|
=
p_i.
\]

Second, suppose that $\hat{\sss}_i$ also belongs to the image of
$T_{\cb'}$. Since $T_{\cb'}$ is injective, there exists a unique index
$j$ such that
$
T_{\cb'}(\sss_j)=\hat{\sss}_i.
$
Thus,
$
Q_{\cb'}(\hat{\sss}_i)=p_j.
$
If $i\in M_{\cb,\cb'}$, then $j\neq i$, because $j=i$ would imply
$
T_{\cb}(\sss_i)=T_{\cb'}(\sss_i),
$
contrary to the definition of $M_{\cb,\cb'}$. By Assumption~\ref{assmp:nonuniform}, the prior probabilities of any two
distinct latent states are separated by at least
\[
\Delta_p
=
\min_{k\neq \ell}|p_k-p_\ell|.
\]
Since $j\neq i$, it follows that
\[
|p_i-p_j|\geq\Delta_p.
\]
Finally, using
\[
Q_{\cb}(\hat{\sss}_i)=p_i,
\qquad
Q_{\cb'}(\hat{\sss}_i)=p_j,
\]
we obtain
\[
|Q_{\cb}(\hat{\sss}_i)-Q_{\cb'}(\hat{\sss}_i)|
=
|p_i-p_j|
\geq
\Delta_p.
\]
Since
$
|Q_{\cb}(\hat{\sss}_i)-Q_{\cb'}(\hat{\sss}_i)|
\geq \Delta_p,
$
we have
$
\frac{
|Q_{\cb}(\hat{\sss}_i)-Q_{\cb'}(\hat{\sss}_i)|
}{
\Delta_p
}
\geq 1.
$
Multiplying by $p_{\max}$ gives
\[
p_{\max}
\leq
\frac{p_{\max}}{\Delta_p}
|Q_{\cb}(\hat{\sss}_i)-Q_{\cb'}(\hat{\sss}_i)|.
\]
Finally, since $p_i\leq p_{\max}$, for every $i\in M_{\cb,\cb'}$,
\[
p_i
\leq
\frac{p_{\max}}{\Delta_p}
|Q_{\cb}(\hat{\sss}_i)-Q_{\cb'}(\hat{\sss}_i)|.
\]

Because $T_{\cb}$ is injective, the points
$\{\hat{\sss}_i\}_{i=1}^K$ are distinct. Summing over
$i\in M_{\cb,\cb'}$ therefore gives
\begin{align*}
P_{\Sb}
\left[
T_{\cb}(\Sb)\neq T_{\cb'}(\Sb)
\right]
&=
\sum_{i\in M_{\cb,\cb'}}p_i\\
&\leq
\frac{p_{\max}}{\Delta_p}
\sum_{i\in M_{\cb,\cb'}}
|Q_{\cb}(\hat{\sss}_i)-Q_{\cb'}(\hat{\sss}_i)|\\
&\leq
\frac{p_{\max}}{\Delta_p}
\sum_{\hat{\sss}}
|Q_{\cb}(\hat{\sss})-Q_{\cb'}(\hat{\sss})|\\
&=
\frac{2p_{\max}}{\Delta_p}
\operatorname{TV}(Q_{\cb},Q_{\cb'}),
\end{align*}
where $\operatorname{TV}(\cdot,\cdot)$ denotes the total variation distance between two probability distributions, defined for discrete distributions $P$ and $Q$ as
\[
\operatorname{TV}(P,Q)
:=
\frac{1}{2}\sum_{z}|P(z)-Q(z)|.
\]

Hence,
\begin{equation}
\label{eq:latent_disagreement_tv}
P_{\Sb}
\left[
T_{\cb}(\Sb)\neq T_{\cb'}(\Sb)
\right]
\leq
\frac{2p_{\max}}{\Delta_p}
\operatorname{TV}(Q_{\cb},Q_{\cb'}).
\end{equation}

By the definition of $D_{\hat S}$,
\[
\|T_{\cb}(\Sb)-T_{\cb'}(\Sb)\|_2^2
\leq
D_{\hat S}^2
\mathbf{1}
\left\{
T_{\cb}(\Sb)\neq T_{\cb'}(\Sb)
\right\}.
\]
Taking expectation over $\Sb$ and applying \eqref{eq:latent_disagreement_tv} yields
\begin{align}
\mathbb E_{\Sb}
\left[
\|T_{\cb}(\Sb)-T_{\cb'}(\Sb)\|_2^2
\right]
&\leq
D_{\hat S}^2
P_{\Sb}
\left[
T_{\cb}(\Sb)\neq T_{\cb'}(\Sb)
\right]
\nonumber\\
&\leq
\frac{2D_{\hat S}^2p_{\max}}{\Delta_p}
\operatorname{TV}(Q_{\cb},Q_{\cb'}).
\label{eq:latent_distance_tv}
\end{align}

\paragraph{Step 3: latent disagreement propagates to conversion error.}
For a realization $\sss$ of $\Sb$, define the source and corresponding
target samples
\[
\xb_0=f(\sss,\cb),
\qquad
\xb'=f(\sss,\cb'),
\]
and the converted output
\[
\hat{\xb}'
:=
d(e(\xb_0),\cb').
\]
The conversion error is therefore
\[
\epsilon_{\mathrm{conv}}(\sss,\cb,\cb')
=
\|\xb'-d(e(\xb_0),\cb')\|_2^2.
\]

By Assumption~\ref{assmp:reconstruction}, applied to the target sample
$\xb'=f(\sss,\cb')$,
\[
\|d(e(\xb'),\cb')-\xb'\|_2^2
\leq\epsilon.
\]
Define the reconstruction residual
\[
\mathbf{r}
:=
\xb'-d(e(\xb'),\cb'),
\qquad
\|\mathbf{r}\|_2^2\leq\epsilon.
\]
Then
\[
\xb'=d(e(\xb'),\cb')+\mathbf{r},
\]
and consequently
\begin{align*}
\epsilon_{\mathrm{conv}}(\sss,\cb,\cb')
&=
\|
d(e(\xb'),\cb')
-
d(e(\xb_0),\cb')
+
\mathbf{r}
\|_2^2\\
&\leq
2
\left(
\|d(e(\xb'),\cb')-d(e(\xb_0),\cb')\|_2^2
+
\|\mathbf{r}\|_2^2
\right)\\
&\leq
2
\left(
\|d(e(\xb'),\cb')-d(e(\xb_0),\cb')\|_2^2
+
\epsilon
\right).
\end{align*}

By Assumption~\ref{assmp:smooth},
\[
\|d(e(\xb'),\cb')-d(e(\xb_0),\cb')\|_2
\leq
L\|e(\xb')-e(\xb_0)\|_2.
\]
Using
\[
e(\xb')=T_{\cb'}(\sss),
\qquad
e(\xb_0)=T_{\cb}(\sss),
\]
we obtain
\[
\epsilon_{\mathrm{conv}}(\sss,\cb,\cb')
\leq
2\left(
L^2
\|T_{\cb'}(\sss)-T_{\cb}(\sss)\|_2^2
+
\epsilon
\right).
\]
Taking expectation over $\Sb$ gives
\begin{align*}
\mathbb E_{\Sb}
\left[
\epsilon_{\mathrm{conv}}(\Sb,\cb,\cb')
\right]
&\leq
2\epsilon
+
2L^2
\mathbb E_{\Sb}
\left[
\|T_{\cb'}(\Sb)-T_{\cb}(\Sb)\|_2^2
\right].
\end{align*}
Applying Eq.~\eqref{eq:latent_distance_tv}, we obtain the intermediate
total-variation bound
\begin{equation}
\label{eq:tv_conversion_bound}
\mathbb E_{\Sb}
\left[
\epsilon_{\mathrm{conv}}(\Sb,\cb,\cb')
\right]
\leq
2\epsilon
+
\frac{4L^2D_{\hat S}^2p_{\max}}{\Delta_p}
\operatorname{TV}(Q_{\cb},Q_{\cb'}).
\end{equation}

\paragraph{Step 4: approximate independence yields the uniform bound.}
Let
\[
Q:=P_{\hat{\Sb}}.
\]
By the triangle inequality for total variation,
\begin{align}
\operatorname{TV}(Q_{\cb},Q_{\cb'})
&\leq
\operatorname{TV}(Q_{\cb},Q)
+
\operatorname{TV}(Q,Q_{\cb'}).
\label{eq:tv_triangle_conditions}
\end{align}
Assumption~\ref{assmp:latentindep} states that
\[
\sup_{\tilde{\cb}\in\operatorname{supp}(\Cb)}
\operatorname{TV}(Q_{\tilde{\cb}},Q)
\leq\delta.
\]
Therefore, for the arbitrary values $\cb,\cb'$ fixed above,
\[
\operatorname{TV}(Q_{\cb},Q)\leq\delta,
\qquad
\operatorname{TV}(Q_{\cb'},Q)\leq\delta,
\]
and Eq.~\eqref{eq:tv_triangle_conditions} gives
\[
\operatorname{TV}(Q_{\cb},Q_{\cb'})
\leq 2\delta.
\]
Substituting this result into
Eq.~\eqref{eq:tv_conversion_bound} yields
\[
\mathbb E_{\Sb}
\left[
\epsilon_{\mathrm{conv}}(\Sb,\cb,\cb')
\right]
\leq
2\epsilon
+
\frac{8L^2D_{\hat S}^2p_{\max}}{\Delta_p}\delta.
\]
Since $\cb$ and $\cb'$ were arbitrary elements of
$\operatorname{supp}(\Cb)$, the bound holds for every admissible
conversion pair, completing the proof.
\end{proof}

\section{Proof of Corollary~\ref{prop:error_bound_discrete}}
\label{app:error_discrete_proof}
\begin{proof}
Let $\hat{\sss},\hat{\sss}'\in\mathbb{R}^K$ be any two categorical
probability vectors. Since their entries are nonnegative and sum to one,
\[
\|\hat{\sss}\|_2^2
=
\sum_{k=1}^{K}\hat{s}_k^2
\leq
\left(\sum_{k=1}^{K}\hat{s}_k\right)^2
=
1,
\]
and similarly,
\[
\|\hat{\sss}'\|_2^2\leq 1.
\]
Moreover, because both vectors are component-wise nonnegative,
\[
\hat{\sss}^{\top}\hat{\sss}'\geq 0.
\]
Therefore,
\begin{align*}
\|\hat{\sss}-\hat{\sss}'\|_2^2
&=
\|\hat{\sss}\|_2^2
+
\|\hat{\sss}'\|_2^2
-
2\hat{\sss}^{\top}\hat{\sss}' \\
&\leq 2.
\end{align*}
Hence, the latent diameter appearing in Proposition~\ref{prop:error_bound}
satisfies
\[
D_{\hat S}^2\leq 2.
\]
Substituting this bound into
\[
\mathbb E_{\Sb}
\left[
\epsilon_{\mathrm{conv}}
\right]
\leq
2\epsilon
+
\frac{8L^2D_{\hat S}^2p_{\max}}{\Delta_p}\delta
\]
gives
\[
\mathbb E_{\Sb}
\left[
\epsilon_{\mathrm{conv}}
\right]
\leq
2\epsilon
+
\frac{16L^2p_{\max}}{\Delta_p}\delta,
\]
which proves the result.
\end{proof}

\section{Continuous Latent Variant}
\label{app:continuous_latent}

For completeness, we also evaluate a continuous variant of the latent representation. In this setting, the encoder outputs
\[
e(x)=\hat{\Sb}\in\mathbb{R}^{T\times D},
\]
using a continuous bottleneck at each time step of dimension $D=1$. This variant follows the same reconstruction and independence training principle as the discrete model. We include it only as an empirical ablation.

Table~\ref{tab:continuos} compares the continuous and discrete variants. The discrete representation achieves slightly better WER, CER, and target-speaker EER.

\begin{table}[t]
    \centering
    \caption{Comparison of the continuous and discrete IVC variants on objective evaluation metrics. The reference speaker is given by a single \textbf{3-second} utterance.}
    \resizebox{0.6\linewidth}{!}{
    \begin{tabular}{lccc}
    \toprule
    Method & WER (\%) $\downarrow$ & CER (\%) $\downarrow$ & EER (\%) $\uparrow$ \\
    \midrule
    Target  & 3.39 & 0.84 & 47.83 \\
    \hline
    \textbf{IVC} $(\text{continuous } \hat{\Sb})$ & 5.30 & 1.60 & 20.50 \\
    \textbf{IVC} $(\text{discrete } \hat{\Sb})$ & 5.29 & 1.59 & \textbf{23.00} \\
    \bottomrule
    \end{tabular}}
    \label{tab:continuos}
\end{table}

\section{Speech Units Analysis}
\label{app:speech_units}

To provide empirical evidence that Assumption~\ref{assmp:nonuniform} holds in real-world speech data, we analyze the variable $\tilde \Sb$ constructed from the $K$-means labels of samples $\{\xb_i\}_{i=1}^N$. We estimate the prior probability of each category by measuring the empirical frequency of the corresponding label in the training dataset,
\[
p_k \;=\; \frac{1}{N}\sum_{i=1}^N \mathbb{I}\{\tilde s_i = k\}, 
\qquad k \in \{1,\dots,K\}.\]
From the resulting probability vector $\mathbf{p} = [p_1, p_2, \dots, p_K]$, we compute the pairwise square-root $l_1$ distance matrix
\(
\mathbf{D} \;=\; \big|\, \mathbf{p}\mathbf{1}^\top - \mathbf{1}\mathbf{p}^\top \,\big|^{\frac{1}{2}},
\)
where $\mathbf{1} \in \mathbb{R}^K$ denotes the all-ones column vector.

\begin{figure}[t]
  \centering
    \includegraphics[width=0.8\linewidth, trim={5 5 5 5},clip]{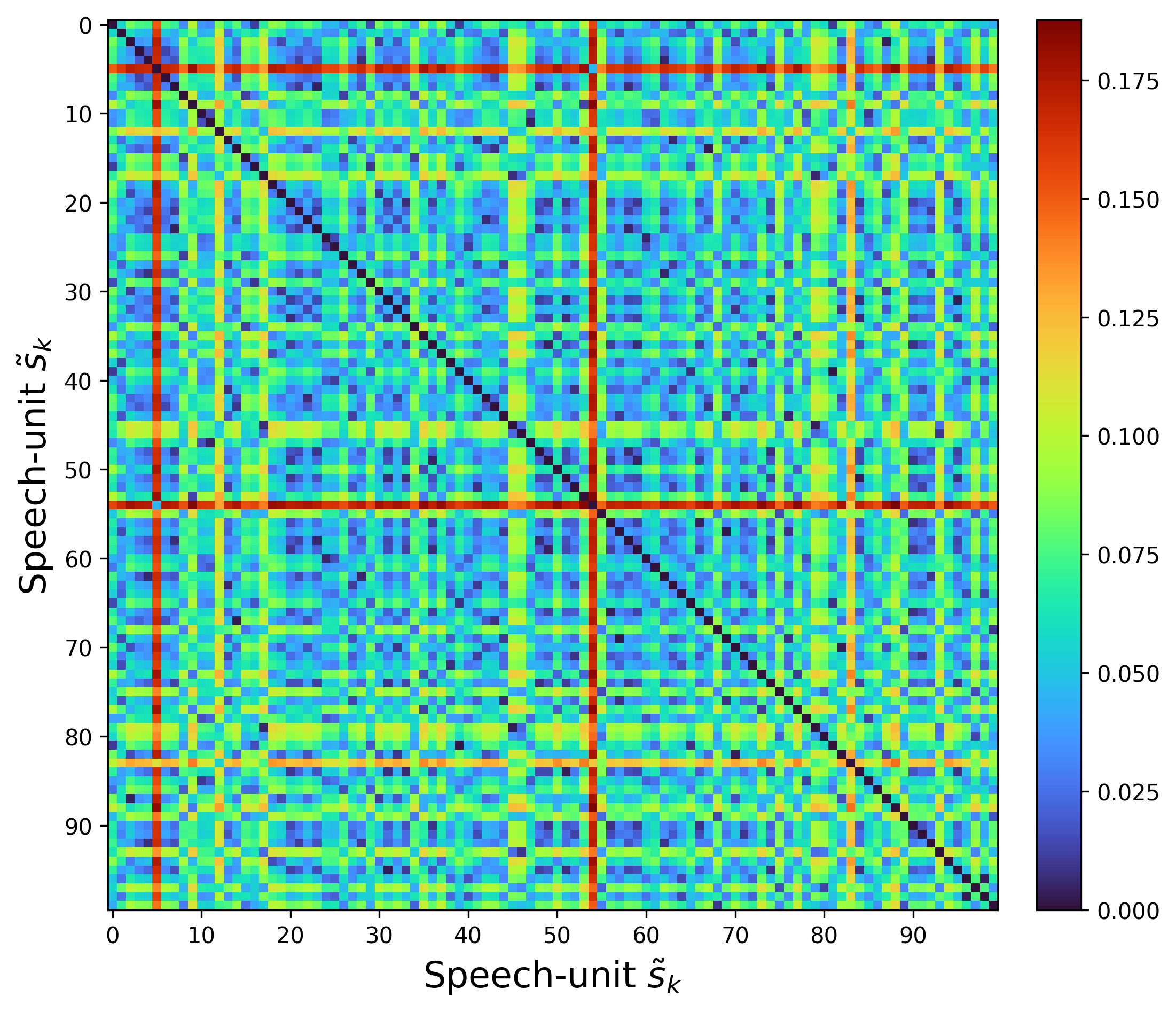}
  \caption{
  Empirical analysis of the asymmetric-prior condition in Assumption~\ref{assmp:nonuniform}. We compute the pairwise distance matrix $\mathbf{D}$ over the empirical prior probabilities $p_k$ of the speech unit categories $\tilde{s}_k$, for $k \in \{1,\dots,K\}$. Since $\tilde{\Sb}$ serves as a proxy for the latent content variable $\Sb$ and $\mathbf{D}_{ii} = 0$ for all $i$ while $\mathbf{D}_{ij} \neq 0$ for all $i \neq j$, the results indicate pairwise-distinct empirical category priors in real-world speech data.
  }
  \label{fig:speeh_units_distance}
\end{figure}

Figure~\ref{fig:speeh_units_distance} visualizes the matrix $\mathbf{D}$. All off-diagonal entries are non-zero and clearly separated from the diagonal, indicating that the empirical prior probabilities of the speech unit categories are distinct. Since the unobserved content variable $\Sb$ is approximated by the proxy $\tilde{\Sb}$, which captures phonetic structure, this analysis provides empirical support for a positive empirical identifiability margin $\Delta_p$ in the speech-unit proxy, while not establishing the assumption for the unobserved true content variable $\Sb$.

\section{Prior Work on Independence Objectives}
Several prior works have explored different strategies to enforce the independence condition. We review them briefly and compare them to our approach.

\textbf{Hilbert-Schmidt Independence Criterion} \hspace{0.1in} \citet{ma2020hsic} proposed to use an empirical estimate of the  Hilbert-Schmidt Independence Criterion (HSIC) \citep{gretton2005measuring} objective, which measures the statistical dependence between two random variables using kernel methods. Hilbert-Schmidt norm of the cross-covariance operator between the distributions in the Reproducing Kernel Hilbert Space (RKHS) defined by
\begin{equation}
\label{eq:app_hsic}
    (N-1)^{-2}\text{tr}(\mathbf{K}_{\hat{\Sb}} \mathbf{H} \mathbf{K}_{\Cb} \mathbf{H}),
\end{equation}
where $\mathbf{K}_{\hat{\Sb}}, \mathbf{K}_{\Cb} \in \mathbb{R}^{N \times N}$ are kernel matrices computed over the sets of samples $\hat{\mathbb{S}} = \{ \hat{\mathbf{s}}_1, \ldots, \hat{\mathbf{s}}_N \}$ and ${\mathbb{C}} = \{ \hat{\mathbf{c}}_1, \ldots, \hat{\mathbf{c}}_N \}$ using a positive-definite kernel function (e.g., Gaussian/RBF), $\mathbf{H} \in  \mathbb{R}^{N \times N}$ is the centering matrix defined by $\mathbf{H}=\mathbf{I}_N-\frac{1}{N}\mathbf{1}_N\mathbf{1}_N^T$ to ensure that the kernels are computed on zero-mean features and $N$ is the number of samples.
Intuitively, HSIC measures the covariance between features in two RKHSs induced by the kernels on $\hat{\Sb}$ and $\Cb$. If the variables are statistically independent, the cross-covariance operator vanishes, and thus the objective in Eq.~\ref{eq:app_hsic} approaches zero. While this biased HSIC estimator is often preferred for its lower variance, it corresponds to a V-statistic and exhibits a finite-sample bias due to the inclusion of self-similarity (diagonal) terms. In contrast, one might employ the unbiased U-statistic estimator, which removes these terms and provides a consistent estimate of statistical dependence \citep{gretton2007kernel}.

\textbf{Contrastive Log-ratio Upper Bound} \hspace{0.1in}
\citet{cheng2020club} proposed a contrastive log-ratio upper bound (vCLUB) of mutual information for a variational autoencoder, which is trained with the following objective function:
\begin{equation*}
    \min_{\thetab} \left[ \mathbb{E}_{\mathbb{P}({\hat{\Sb}},\Cb)}[\log q_\thetab(\Cb|{\hat{\Sb}})]  - \mathbb{E}_{\mathbb{P}({\hat{\Sb}})}\mathbb{E}_{\mathbb{P}(\Cb)}[\log q_\thetab(\Cb | {\hat{\Sb}})] \right],
\end{equation*}
where $q_\thetab(\Cb|{\hat{\Sb}})$ is a variational distribution parametrized by $\thetab$ that approximates $\mathbb{P}(\Cb|{\hat{\Sb}})$. The intuition behind the vCLUB objective is to estimate the mutual information between two variables by contrasting how well a variational model can predict the true paired samples versus mismatched (independent) samples. The first term encourages a high likelihood of true pairs (\(\Sb,\Cb\)). In contrast, the second term penalizes a high likelihood of randomly paired \(\Sb\) and \(\Cb\), sampled independently from their marginals. The difference between these two expectations provides an upper bound on the mutual information, which can be minimized to encourage statistical independence between the variables.

\textbf{Adversarial Independence} \hspace{0.1in}
Another approach was proposed by \citet{shaham2022discovery}
where the model is trained in an adversarial way, and a discriminator $g(e(\mathbf{x}))$ is trained to predict a condition $\mathbf{c}$ from the latent $\hat{\mathbf{s}}=e(\mathbf{x})$ by maximizing the objective $-\mathcal{I}(g({\hat{\mathbf{s}}}), \mathbf{c})$, e.g. minimizing cross entropy loss term $ \sum_{i=1}^{K} c_i \log(g(\hat{\mathbf{s_i}}))$ between condition $\mathbf{c}$ and discriminator prediction $g(\hat{\mathbf{s}})$. The autoencoder is trained to confuse the discriminator by minimizing  $-\mathcal{I}(g(e(\mathbf{x})), \mathbf{c})$. Assuming a problem with a single conditional source $\Cb$, the training objective becomes:
\begin{equation*}
    \min_{e, d} \max_g [ \mathcal{R} (\mathbf{x}, d({\hat{\mathbf{s}}}, \mathbf{c})) - \lambda \mathcal{I} (g({\hat{\mathbf{s}}}), \mathbf{c})].
\end{equation*}
While this approach has been proven to recover the target latent component up to an entropy-preserving transformation, it critically relies on the capacity and stability of the discriminator. In practice, weak or poorly trained discriminators may suffer from mode collapse, where the discriminator focuses only on a subset of easily distinguishable modes in the conditional variable and ignores others. As a result, the encoder may exploit this weakness by only obfuscating the modes to which the discriminator is sensitive, while still leaking conditional information through other dimensions. This undermines the goal of achieving valid conditional invariance and can lead to incomplete or biased disentanglement in the learned representation.

\section{Limitations}
\begin{itemize}
    \item \textbf{Model Architecture} \hspace{0.15in} 
    The proposed method relies on the quality and expressiveness of the pretrained self-supervised learning (SSL) encoder and acoustic decoder. Since our autoencoder operates on the outputs of the SSL encoder and its reconstructions serve as inputs to the acoustic decoder, any limitations or biases in these components can affect the performance and fidelity of the conversion. However, this dependence also becomes a strength in low-resource scenarios: the SSL and acoustic models are foundational models trained on large-scale, diverse datasets, enabling strong generalization even when the trainable part of our method is relatively lightweight. As a result, our approach remains effective and data-efficient in domains with limited labeled or supervised data.

    \item \textbf{Evaluation and Baselines} \hspace{0.15in}  This work prioritizes the theoretical analysis and convergence guarantees of our proposed framework over achieving state-of-the-art (SOTA) empirical performance. Due to limited research resources and computational constraints, we train and evaluate our method on publicly available datasets with reduced scale. For fair comparison, we benchmark against baselines trained under the same conditions. While some recent models (e.g., NANSY, SelfVC) report strong empirical results, they were trained on large-scale or proprietary datasets and do not publicly release complete code or training details, making direct comparison infeasible. Our goal is to provide a principled and reproducible foundation that can support future extensions and scaling efforts.
\end{itemize}

\section{Broader Impacts}
This work presents a model for speech attributes conversion, e.g. voice or emotion, offering benefits such as improved accessibility, expressive speech synthesis, and enhanced human-computer interaction. However, it also poses risks, including potential misuse for impersonation, emotional manipulation, and audio deepfakes. These concerns are particularly relevant to disinformation and privacy violations. Our work is intended for controlled research use, and we emphasize the need for future safeguards, such as watermarking, detection tools, and responsible access policies, to mitigate misuse and uphold ethical standards.

\section{Additional Results}
\begin{figure}[h]
\centering
    \centering
    \includegraphics[width=0.6\linewidth]{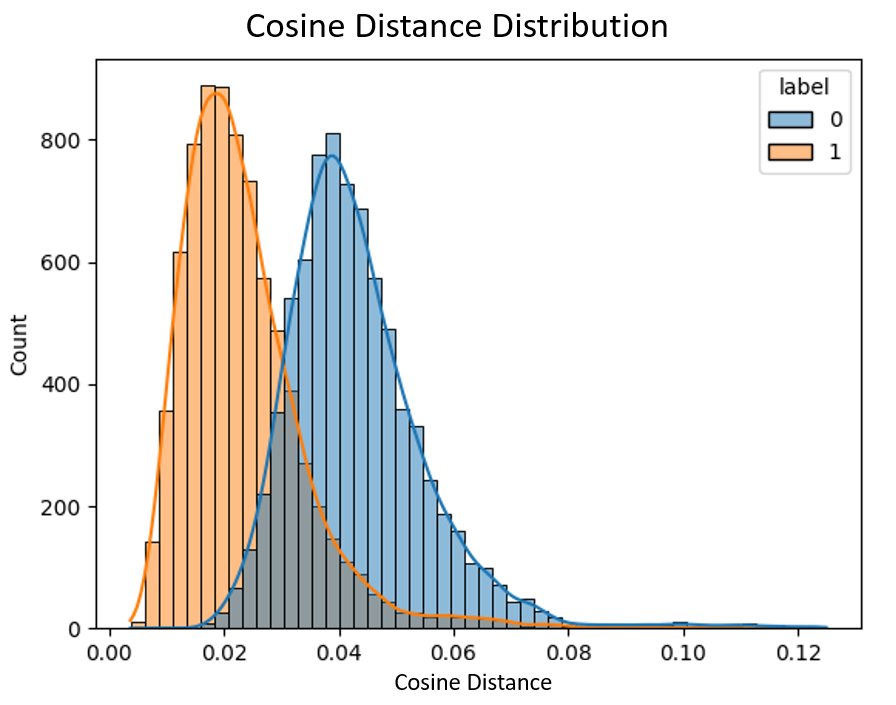}
    \caption{Distributions of cosine distances calculated on the pairs constructed from real samples (orange) and on the pairs constructed such that sample one comes from generated samples and sample two comes from a real sample (blue). The distance is defined by $1-\frac{\mathbf{x}_i \cdot \mathbf{x}_j}{||\mathbf{x}_i||_2||\mathbf{x}_j||_2}$ where $\mathbf{x}_i$, $\mathbf{x}_j$ are speaker embeddings. The results obtained from the model trained with the RedimNet speaker encoder. The plot shows that the generated samples are very close to the real samples and almost indistinguishable.}
    \label{fig:cos_dists}
\end{figure}

The cosine distance between the generated and real samples is very low, indicating that the generated samples closely match real speaker characteristics. We demonstrate this by plotting the distribution of cosine distances between real and generated samples in Figure \ref{fig:cos_dists}.

\section{Training Details}
Table~\ref{tab:training_details} summarizes the hyperparameters used for training. The model is trained on source audio segments of 3 seconds in length, ensuring that all features are extracted from short-duration waveforms. All models were trained using the Adam optimizer to reduce the memory demands of the moments.

\begin{table}[h]
\centering
\caption{Training hyperparameters}
\label{tab:training_details}
\begin{tabular}{lc}
\hline
   Hyperparameter             & Value  \\
\hline
Number of parameters & 21.4M \\
Epochs & 100 \\
Batch Size & 256 \\
LR (start)  & 1e-3 \\
LR (end)  & 1e-5 \\
Scheduler & cosine \\
Segment Length (sec) & 3\\
Condition Encoder & RedimNet \\
Optimizer & Adam \\
\hline
\end{tabular}

\end{table}

\section{Mel-Spectrogram IVC Variant with Vocos}
\label{app:mel_ae_vocos}

We also evaluate an acoustic-domain variant of IVC that removes the pretrained WavLM feature extractor. 
Instead, the observed variable $\Xb$ is defined as the mel-spectrogram of the input waveform. 
The autoencoder encodes the mel-spectrogram into $\hat{\Sb}=e(\Xb)$ and reconstructs it with the decoder conditioned on the target attribute $\Cb$, using the same conditioning setup as in the main experiments.

Waveform synthesis is performed with Vocos. 
Training is done in two stages. 
First, the mel-spectrogram autoencoder is trained with the reconstruction objective and independence penalty between $\hat{\Sb}$ and $\Cb$ as in Eq.~\ref{eq:general_loss}. 
Second, the autoencoder is frozen and Vocos is fine-tuned on converted mel-spectrograms produced by the frozen model. 
This adapts the vocoder to the distribution of converted rather than natural mel-spectrograms. 
Notably, our model contains only 20.5M parameters, making it a compact alternative to SSL-based conversion systems.

This variant serves as an ablation of the representation choice: it tests whether IVC can operate directly on mel-spectrograms without SSL speech features. 
We compare against MeanVC, a recently proposed lightweight voice-conversion method in Table~\ref{tab:mel_ae_vocos_results}.

\begin{table}[t]
    \centering
    \caption{Objective evaluation of the mel-spectrogram autoencoder variant with Vocos vocoder.}
    \label{tab:mel_ae_vocos_results}
    \resizebox{0.7\linewidth}{!}{
    \begin{tabular}{lc|ccc}
        \toprule
        Method & Params (M)  & EER (\%) $\uparrow$ & WER (\%) $\downarrow$ & CER (\%) $\downarrow$ \\
        \midrule
        IVC-mel & 20.5  & \textbf{11.05} & 29.40 & 13.77 \\
        MeanVC & 14.0 & 6.14 & \textbf{29.26} & \textbf{13.25}  \\
        \bottomrule
    \end{tabular}}
\end{table}

\end{document}